\documentclass[prl,showpacs,twocolumn]{revtex4}
\usepackage{graphicx}
\usepackage{amsmath}
\usepackage{amsfonts}
\begin{document}

\title{Superflow in a toroidal Bose-Einstein condensate:\\ an atom circuit with a tunable weak link}

\author{A. Ramanathan$^*$}
\affiliation{Joint Quantum Institute, National Institute of Standards and Technology and University of Maryland, Gaithersburg, MD, 20899, USA}

\author{K. C. Wright}\email[The first two authors contributed equally. Correspondence may be directed to: ]{ kcwright@nist.gov}
\affiliation{Joint Quantum Institute, National Institute of Standards and Technology and University of Maryland, Gaithersburg, MD, 20899, USA}

\author{S. R. Muniz}
\altaffiliation[Current Address: ]{Instituto de F\'{i}sica de S\~{a}o Carlos, Universidade de S\~{a}o Paulo, S\~{a}o Carlos, 13560-970, Brazil}

\author{M. Zelan}
\altaffiliation[Primary Address: ]{Department of Physics, Ume\aa~University, SE-901 87, Ume\aa, Sweden}

\author{W.~T.~Hill~III}
\affiliation{Joint Quantum Institute, National Institute of Standards and Technology and University of Maryland, Gaithersburg, MD, 20899, USA}

\author{C. J. Lobb}
\affiliation{Joint Quantum Institute, National Institute of Standards and Technology and University of Maryland, Gaithersburg, MD, 20899, USA}

\author{K. Helmerson}
\altaffiliation[Current Address: ]{School of Physics, Monash University, 3800, Australia}

\author{W. D. Phillips}
\affiliation{Joint Quantum Institute, National Institute of Standards and Technology and University of Maryland, Gaithersburg, MD, 20899, USA}

\author{G. K. Campbell}
\affiliation{Joint Quantum Institute, National Institute of Standards and Technology and University of Maryland, Gaithersburg, MD, 20899, USA}

\date{\today}

\begin{abstract}
We have created a long-lived ($\approx$ 40 s) persistent current in a toroidal Bose-Einstein condensate held in an all-optical trap. A repulsive optical barrier across one side of the torus creates a tunable weak link in the condensate circuit, which can affect the current around the loop. Superflow stops abruptly at a barrier strength such that the local flow velocity at the barrier exceeds a critical velocity. The measured critical velocity is consistent with dissipation due to the creation of vortex-antivortex pairs. This system is the first realization of an elementary closed-loop atom circuit.
\end{abstract}

\pacs{03.75.Lm, 03.75.Kk, 67.85.De}

\maketitle

Quantum fluids can exhibit properties such as long-range coherence and superfluidity that make them useful for constructing sensors and other devices. For example, superconducting quantum interference devices (SQUIDs) are sensitive magnetic field detectors~\cite{ClarkeSQUID04}, and superfluid He circuits have been used to detect rotation~\cite{SimmondsQuantumN01,HoskinsonSuperfluidPRB06}. Ultracold atomic-gas analogs of electronic devices and circuits, or ``atomtronics'' have been proposed including diodes and transistors~\cite{atomtronics}. Of particular interest is the realization of an atomic-gas SQUID analog.  SQUID circuits have been realized with either tunnel or weak link junctions ~\cite{ClarkeSQUID04, LikharevRMP79, DavisRMP02}. In atomic Bose-Einstein condensates, Josephson junctions have been demonstrated only between adjacent wells~\cite{AlbiezDirectPRL05, LevyACDCN07}. Here we present the first implementation of a non-trivial, closed-loop atom circuit, and show that it is possible to control the current at the single-quantum level by changing the strength of a weak link. This is an essential step toward realizing an atomic SQUID analog.

Superfluids flow without dissipation if the flow velocity is below a threshold determined by the lowest energy excitations allowed for the system~\cite{LandauJPUSSR41}. In a homogeneous condensate the lowest energy excitations are phonons~\cite{endnote}, and the Landau critical velocity is the speed of sound~\cite{PitaevskiiBose-Einstein03}. Real systems are finite and therefore inhomogeneous; consequently the lowest energy excitations can be vortex-like~\cite{BarenghiQuantized01}, and dissipation can occur at velocities well below the sound speed~\cite{FeynmanApplicationPLTP55}. Dissipation involving vortex-like excitations has been previously observed in experiments with liquid He~\cite{AvenelObservationPRL85, AmarQuantizedPRL92}, superconductors~\cite{HuebenerMagnetic01}, and in a simply-connected condensate~\cite{NeelyObservationPRL10}.

\begin{figure}
\centering
\includegraphics[width=0.48\textwidth]{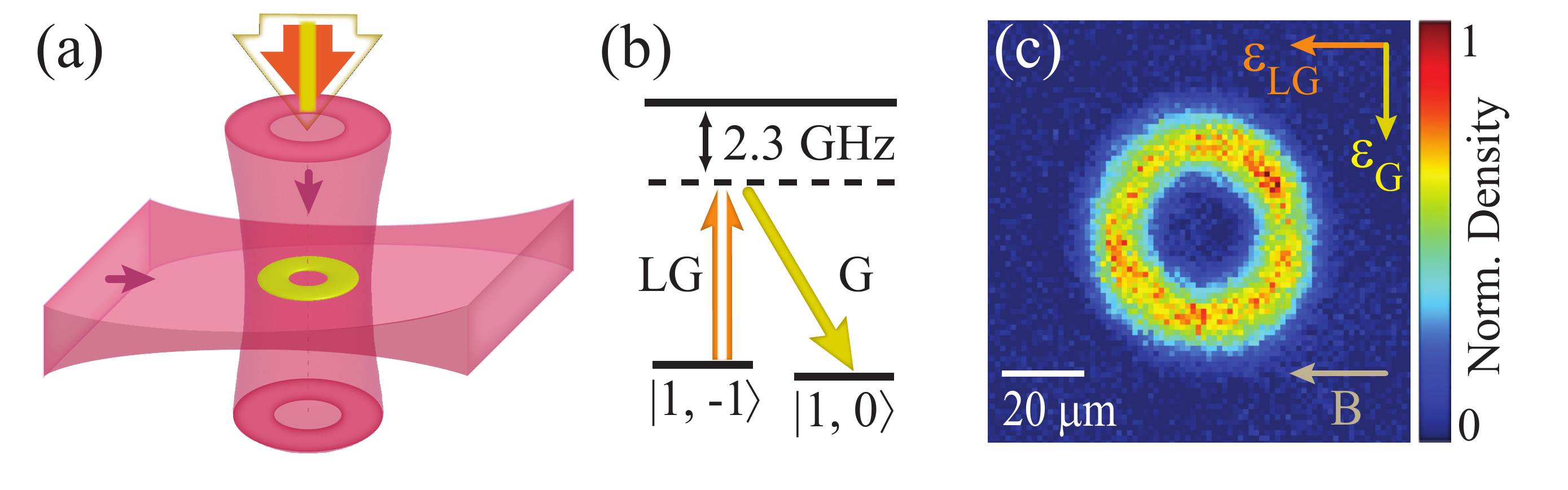}
\caption{Experimental configuration. 
(a) Schematic of the toroidal optical dipole trap formed at the intersection of two red-detuned beams: a horizontal ``sheet'' beam, and a vertical Laguerre-Gaussian beam (LG$_0^1$) with a ring-shaped intensity maximum.  A pulsed pair of Raman beams (large downward arrows) co-propagating with the LG trapping beam creates circulation in the condensate. 
(b) Energy level diagram for the Raman transition: $|F\!=\!1,m_F\!=\!-1\rangle\!\rightarrow\!|F\!=\!1,m_F\!=\!0\rangle$. One Raman beam carries $\hbar$ orbital angular momentum per photon (LG$_0^1$), the other carries none (Gaussian); the condensate is transferred to a quantized ($l$=1) circulating state. 
(c) False-color absorption image showing the normalized column density of a condensate in the trap, viewed from above. Arrows: Raman beam polarizations and magnetic bias. }
\label{fig1}
\end{figure}

The critical velocity in simply-connected condensates has been measured previously by moving a defect created by a localized optical potential~\cite{ OnofrioObservationPRL00, EngelsStationaryPRL07, NeelyObservationPRL10}. When the velocity of the defect was high enough, excitations and heating were observed. In contrast to this earlier work, we create a quantized, persistent flow around a multiply-connected (toroidal) condensate, and study the decay of that flow in the presence of a stationary barrier, as a function of barrier height and condensate atom number. 

In previous experiments~\cite{RyuObservationPRL07}, we created persistent currents in a harmonic magnetic potential pierced by a repulsive optical potential. Relative drift between these potentials limited the flow lifetime to $\approx$ 10 s. This motivated the construction of an all-optical trap which supports persistent currents for up to 40 s, and allows us to carefully study the stability of superflow.

To create a toroidal condensate, $3^2S_{1/2}\,|F\!=\!1,m_F\!=\!-1\rangle\ $ $^{23}$Na atoms are cooled almost to degeneracy in a magnetic trap and then transferred into an optical dipole trap created by the intersection of red-detuned (1030 nm) ``sheet'' and ``ring'' beams (Fig.~\ref{fig1}a). The horizontal sheet beam has a vertical (horizontal) $1/e^2$ half-width of $\approx$ 9 $\mu$m ($\approx$ 400 $\mu$m), and provides vertical confinement. The vertical ring beam is Laguerre-Gaussian (LG$^1_0$), and confines the condensate to its $\approx$ 20 $\mu$m radial intensity peak, generating a toroidal potential minimum. With the atoms in the optical trap, the beam intensities are ramped down to force evaporative cooling. At the end of the ramp, the trap depth is $\approx$ 700 nK, with trap frequencies  $\omega_z/2\pi=550$~Hz (vertical) and $\omega_r/2\pi=110$~Hz (radial). This produces a toroidal condensate of up to 3$\times$10$^5$ atoms with a chemical potential $\mu_0$ of up to $h\cdot1200$ Hz, and temperature $<$~10 nK (no discernible non-condensed fraction). The azimuthal variation of the potential minimum is less than $h\cdot100$ Hz, as shown by the smooth condensate density profile in Fig.~\ref{fig1}c. 
 
  The condensate is initially nonrotating~\cite{yeswearesure}. Superfluid circulation around any closed path must be quantized, such that the wave function has a $2\pi l $ phase winding ($l\!\in\!\mathbb{Z}$). We create circulation by transferring quantized angular momentum from optical fields during a Raman process~\cite{WrightOpticalPRA08}. The co-propagating Raman beams, detuned 2.3 GHz below the D$_2$ transitions, are in two-photon resonance with the $|1,-1\rangle\!\rightarrow\!|1,\;0\rangle\ $ transition (Fig.~\ref{fig1}b). They have orthogonal linear polarizations, parallel and perpendicular to the horizontal magnetic bias field (Fig.~\ref{fig1}c). The nonlinear Zeeman shift from the 0.5 mT field applied during the interaction prevents coupling to $|1,1\rangle$.
  
   The angular momentum change of the condensate is determined by the spatial mode of the Raman beams. With one beam Gaussian, and the other in an LG$^1_0$ spatial mode carrying $\hbar$ orbital angular momentum per photon, the Raman process coherently transfers the condensate to the $l$=1 circulating state~\cite{AndersenQuantizedPRL06, WrightOpticalPRA08, RyuObservationPRL07}. With good mode matching and an optimized Raman $\pi$-pulse ($\approx$ 100 $\mu$s), we achieve a minimum transfer efficiency of  90\%, with only a few percent atom loss due to spontaneous scattering. Residual atoms in $|1,-1\rangle$ after the Raman pulse are quickly removed from the trap by transferring them to $|2,-2\rangle$ with a microwave pulse, then ejecting them from the trap with resonant imaging light (see below).
  
\begin{figure}
  \centering
\includegraphics[width=0.46\textwidth]{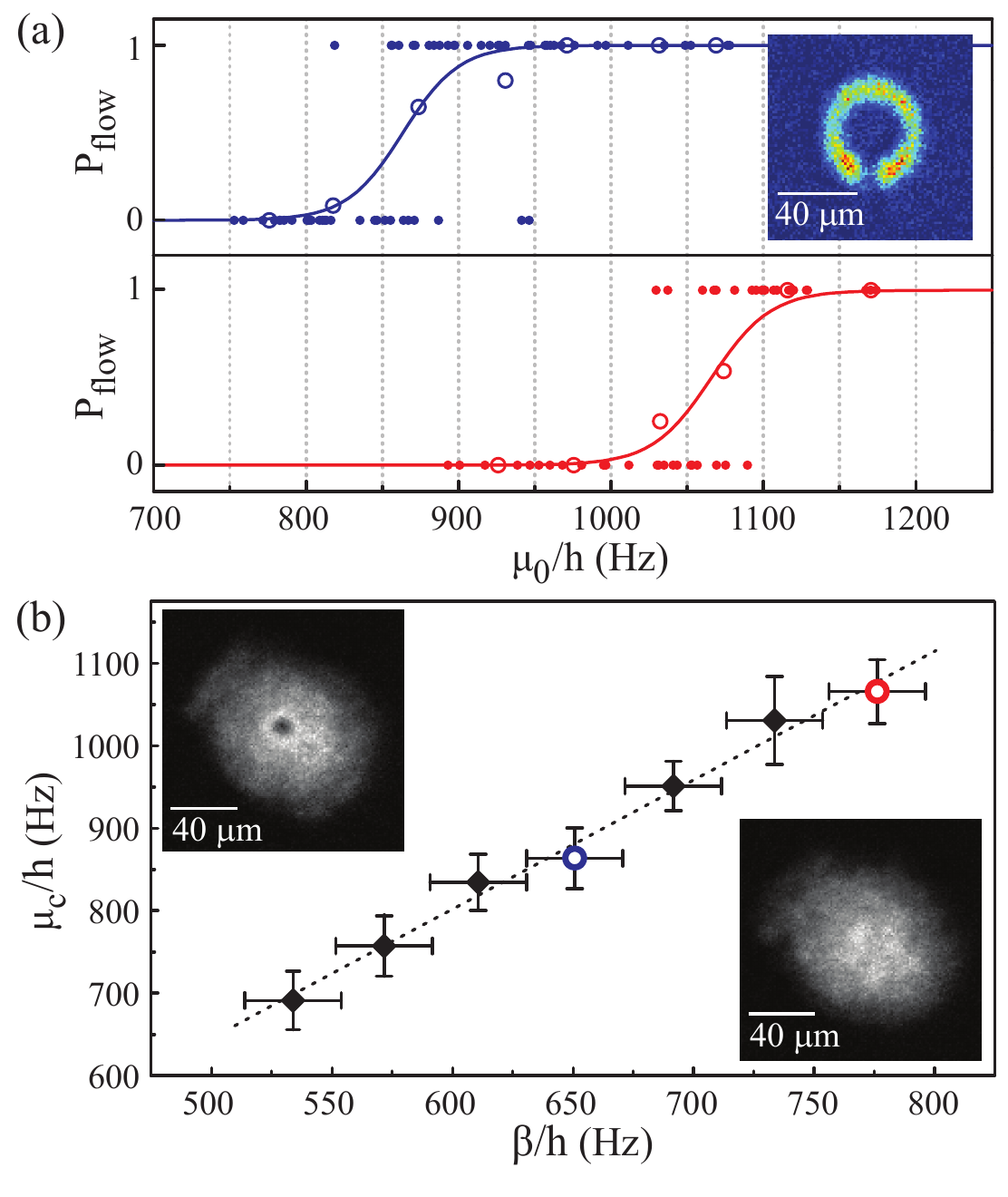}
  \caption {(a) Flow survival as a function of chemical potential, $\mu_0$, for two barrier heights: $\beta/h$ = 650 Hz (upper, blue), and $\beta/h$ = 780 Hz (lower, red). Presence or absence of flow for a single condensate is shown by closed circles. Open circles are the average of data within the bins (vertical lines), representing the flow survival probability (P$_\textrm{flow}$) of each bin. A critical chemical potential $\mu_c$ for stable flow is found from a sigmoidal fit (solid lines) to the data for each $\beta$. Inset: \textit{In situ} absorption image of a condensate near $\mu_c$ ($\mu_0/h$ = 870 Hz, $\beta/h$ = 650 Hz). (b) Values of $\mu_c$ at different $\beta$, determined by fits as described in (a). The open circles correspond to the data in (a). The vertical error bars reflect the width of the sigmoidal fit, $\pm 2\mu_w$. The horizontal error bars are the 20 Hz uncertainty in calibrating $\beta$. The dotted line is a linear fit to the data, with slope 1.6(2). Insets: typical TOF absorption images showing the presence (top left) and absence (bottom right) of circulation.}
  \label{SurvivalProb}
\end{figure}

Circulation is detected by releasing the condensate from the trap and imaging the density distribution after several milliseconds time-of-flight (TOF)~\cite{MadisonVortexPRL00}. If the condensate is not rotating, the central hole closes after a short time.  When a rotating condensate is released, the angular velocity of the flow prevents complete closure. The persistence of a central hole after sufficiently long TOF is the signature of circulation in the ring (see Fig~\ref{SurvivalProb}(b) insets). The apparent size of the hole at a given time after release is related to the the azimuthal flow velocity prior to release and the velocity of the mean-field-driven inward expansion. For a rotating condensate released directly from our narrow annular trap, the hole size is below our imaging resolution for experimentally accessible TOFs ($<$ 15 ms). To make the signature of circulation visible earlier, we first adiabatically reduce the ring beam intensity by 90\% over 100 ms, then release the condensate suddenly ($<$ 1 $\mu$s). We use this procedure, followed by 6 ms TOF, to detect circulation. 
  
The Raman beams used to create circulation also cause small-amplitude oscillations in the radial density profile, due to small dipole forces and atom loss. These oscillations have no observable impact on the stability of the circulation, and damp out after $<0.5$~s. We add a wait time $\geq$ 3 s after the Raman transfer to ensure complete damping. The circulation is extremely robust, and continues until losses due to background collisions reduce $\mu_0$ to the level of the nonuniformities in the trap. For a 30 s vacuum-limited $1/e$ condensate lifetime, $\mu_0$ remains high enough for flow to survive up to 40 s.

After the $\geq$ 3 s wait time, we insert a barrier into the path of the flow and study the stability of the circulation. The repulsive barrier is created with a blue-detuned (532 nm) laser beam focused to an elliptical spot. The major axis (15 $\mu$m $1/e^2$ radius) is aligned in the radial direction of the toroid, and the minor axis (4.3 $\mu$m $1/e^2$ radius) is parallel to the flow, and exceeds the bulk condensate healing length ($\xi=\hbar / \sqrt{2 m \mu}$ $<$ 1 $\mu$m). The barrier depletes the local density of the condensate, $n$, as seen in the inset in Fig.~\ref{SurvivalProb}(a). The reduction in density increases the local flow velocity (roughly $v\!\propto\!1/n$). Lowering the density also lowers the local interaction energy, $\mu_l\!\propto\!n$, decreasing the local sound speed. To study flow stability, we ramp up the barrier intensity over 100 ms to a chosen barrier height $\beta$, hold for 2 s, then ramp back down in another 100 ms. The presence or absence of flow is then detected in TOF as described above. This procedure is repeated many times for the same $\beta$, varying the total number of atoms (by varying the initial condensate number and/or the wait time) until the range of atom number is well-sampled. We then change $\beta$ and repeat the procedure. If the barrier is not applied, the flow always survives, so we can attribute the decay of the flow to the effect of the barrier. Separate measurements indicate that the flow decays in $<$ 100 ms.

The analysis of flow stability depends on \textit{in situ} observations of the condensate density profile in the presence of the barrier, and from TOF images after the barrier has been removed.  From TOF images we determine whether the flow survived [insets Fig.~\ref{SurvivalProb}(b)], and measure the condensate atom number, $N$. For an annular condensate with a Thomas-Fermi profile,
the chemical potential $\mu_0 = \hbar\bar{\omega}\sqrt{\pi/2\cdot(N a_s/R)}$ where $\bar{\omega}\equiv\sqrt{\omega_z\omega_r}$, $a_s$ is the s-wave scattering length, and $R$ is the radius of the ring. This calculation does not include small corrections ($\approx 6\%$) due to the azimuthal nonuniformity of the potential minimum and displacement of atoms from the barrier region, corrections which are less than the systematic uncertainty in determining $\mu_0$ ($\approx 10\%$). 

We calibrate $\beta$ by taking \textit{in situ} images of the condensate and measuring the reduction in column density at the location of the barrier (see Fig.~\ref{SurvivalProb}a inset). Due to the high optical depth (up to 10), we use a partial transfer imaging technique~\cite{FreilichReal-TimeS10, RamanathanTBP}, in which a precise fraction (ranging from 15-40\%) of the atoms is transferred to the $|2,-1\rangle$ state using a microwave pulse, then resonantly imaged on the $S_{1/2}\,F\!=\!2\rightarrow P_{3/2}\,F\!=\!3$ transition. The local interaction energy $\mu_l$ can be found from the measured column density $\tilde{n}$.  For data where $\mu_l\!<\!\hbar\omega_z$, we assume the axial density profile is that of the harmonic oscillator ground state, with 
$\mu_l\!=\![8\pi\cdot (\hbar\omega_z) (\hbar^2 a_s^2\tilde{n}^2/m)]^{1/2}$.
For data where $\mu_l>\hbar\omega_z$, we assume a Thomas-Fermi profile, with 
$ \mu_l\!=\![9\pi^2/2\cdot (\hbar\omega_z)^2 (\hbar^2 a_s^2\tilde{n}^2/m)]^{1/3}$.
When measuring the column density at the barrier, we correct for loss of contrast due to the imaging resolution, which reduces the apparent depth of the density depletion by $\approx$ 15\%.  We take $\beta$ to be $\mu_0-\mu_l$.

\begin{figure}
\centering
\includegraphics[width=0.48\textwidth]{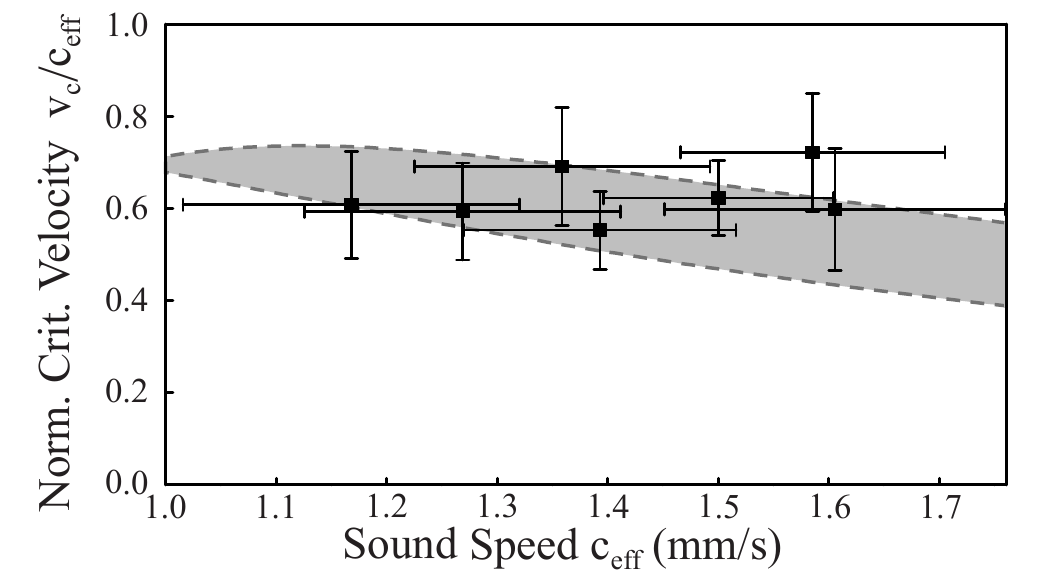}
\caption 
{Critical flow velocity, $v_c$, above which the circulation becomes unstable, for each of the barrier heights in Fig~\ref{SurvivalProb}(b).  The value of $v_c$ is shown normalized to the effective sound speed at the barrier ($c_{\mathrm{eff}}$), and is plotted as a function of $c_{\mathrm{eff}}$.  The horizontal error bars are the estimated uncertainty in $c_{\mathrm{eff}}$. The vertical error bars are the combined experimental uncertainty in $v_c$ and $c_{\mathrm{eff}}$. The measured ratio $v_c/c_{\mathrm{eff}}\approx 0.6$, and is independent of $c_{\mathrm{eff}}$ to within the experimental uncertainty. Vortex-like excitations are expected to occur in this system at and above a velocity $v_F$, where $v_F < c_{\mathrm{eff}}$~\cite{FeynmanApplicationPLTP55}. The gray band indicates estimated upper/lower bounds (see text) on $v_F/c_{\mathrm{eff}}$, using Feynman's approximate expression for $v_F$.}
\label{FlowVelocity}
\end{figure}  

 The solid circles in Fig.~\ref{SurvivalProb}(a) show the flow survival or decay for single experimental runs, plotted against $\mu_0$ for that run. The open circles are the average of the solid circles within the bins shown. The upper plot (blue) is for $\beta/h$ = 650 Hz (upper); the lower (red) shows $\beta/h$ = 780 Hz. At low $\mu_0$, the flow is arrested by the barrier. At high $\mu_0$ the flow survival probability becomes unity. In between, the survival probability increases from zero to one over a narrow critical region. We characterize this critical region for each $\beta$ by fitting a sigmoidal function $P(\mu_0)=1/(1+e^{(\mu_c-\mu_0)/\mu_w})$ to each \textit{unbinned} data set, where the parameters $\mu_c$ and $\mu_w$ are the critical chemical potential and the sigmoidal width respectively~\cite{uncertaintynote}. The observed width is consistent with observed shot-to-shot variations in the trapping potential.
 
 Figure~\ref{SurvivalProb}(b) shows the values of $\mu_c$ extracted from the fits of the  data for seven different $\beta$. Over this range, $\mu_c$ increases approximately linearly with $\beta$, with a slope greater than unity. The functional dependence and slope are determined in a non-trivial way by trap geometry and the condition of quantized circulation around the ring. The experimental results are consistent with expectations for our geometry.

The physics behind Fig.~\ref{SurvivalProb} is more apparent when the data is recast in terms of flow velocity and sound speed at the barrier. The barrier thickness is greater than $\xi$, so we expect the flow to become unstable when the velocity in the barrier region exceeds some local critical velocity $v_c$. The flow velocity at the barrier cannot be determined just from $\mu_0$ and $\beta$. The requirements of quantized circulation (global), and flow conservation (local), make it necessary to self-consistently calculate the velocity distribution around the entire ring. We do this by integrating the \textit{in situ} column density radially to make a 1D approximation of the density profile, then solving for the velocity distribution of an $l=1$ circulation state.

The critical velocity is determined by the lowest energy excitations allowed for the system~\cite{LandauJPUSSR41}. For phonon-like excitations in the ring, that velocity should be approximately the local sound speed in the barrier region~\cite{WatanabeCriticalPRA09}. We make an initial estimate for the critical velocity from the local interaction energy at the peak of the barrier, $c_l=\sqrt{\mu_l/m}$. However, the inhomogeneous (nearly parabolic) radial density profile lowers the effective sound speed to $c_{\mathrm{eff}} = c_l/\sqrt{2}$ for waves traveling azimuthally along the annulus~\cite{StringariDynamicsPRA98}. Figure~\ref{FlowVelocity} shows the observed critical velocity normalized to $c_{\mathrm{eff}}$, as a function of $c_{\mathrm{eff}}$. As seen in previous work with finite inhomogeneous atomic condensates~\cite{OnofrioObservationPRL00, EngelsStationaryPRL07, NeelyObservationPRL10}, the observed critical velocity is less than the sound speed. For all tested values of $\beta$,  $v_c/c_{\mathrm{eff}}\approx 0.6$ and is independent of $c_{\mathrm{eff}}$ to within the experimental uncertainty.

In this experiment, flow is confined to a narrow, flattened channel, raising the possibility that vortex-like excitations are responsible for the observed critical velocity. Numerical simulations~\cite{PiazzaVortex-inducedPRA09} with a model condensate similar to ours, but in an $l=8$ circulation state, showed vortices traversing the barrier region when the barrier was raised above a critical level. This suggests that for our $l=1$ circulation state, a similar decay mechanism could be at work. For vortex-like excitations in our quasi-2D geometry, the (Feynman) critical velocity $v_F$ can be estimated from energetic arguments~\cite{FeynmanApplicationPLTP55} to be $v_F= (\hbar/md)\ln(d/a)$, where $d$ is the channel width, and $a$ is the vortex core size. We take $d$ to be the Thomas-Fermi width, and $a$ the healing length, both calculated for the barrier region. Both $d$ and $a$ depend on $c_{\mathrm{eff}}$ via the interaction energy $\mu_l$. The grey band in Fig.~\ref{FlowVelocity} is an estimate of the probable value of $v_F/ c_{\mathrm{eff}}$ with $c_l\geq c_{\mathrm{eff}}\geq c_l/\sqrt{2}$. While this calculation is in surprisingly good agreement with our data, a more complete model including geometric factors is needed to accurately calculate the energy of a vortex-antivortex pair in the barrier region.

We have presented the first realization of a closed atomtronic circuit, demonstrating precise control both in inducing and arresting superfluid flow. We have clearly identified the critical velocity where flow stops, and our observations are in agreement with theoretical predictions in which vortex-antivortex excitations are the decay mechanism for the system. In future work, we plan to investigate the role of barrier geometry, condensate temperature, and dimensionality in determining the critical velocity and decay mode. In addition, rotating a barrier around the ring (oscillating it azimuthally) would be analogous to magnetically biasing (driving an AC current in) a SQUID. The present work constitutes a significant step toward realizing such an atomic SQUID analog.

The authors thank L. Mathey for helpful discussions, and R. B. Blakestad for comments on the manuscript. This work was partially supported by ONR, the ARO atomtronics MURI, and the NSF PFC at JQI.


\begin{thebibliography}{30}
\expandafter\ifx\csname natexlab\endcsname\relax\def\natexlab#1{#1}\fi
\expandafter\ifx\csname bibnamefont\endcsname\relax
  \def\bibnamefont#1{#1}\fi
\expandafter\ifx\csname bibfnamefont\endcsname\relax
  \def\bibfnamefont#1{#1}\fi
\expandafter\ifx\csname citenamefont\endcsname\relax
  \def\citenamefont#1{#1}\fi
\expandafter\ifx\csname url\endcsname\relax
  \def\url#1{\texttt{#1}}\fi
\expandafter\ifx\csname urlprefix\endcsname\relax\def\urlprefix{URL }\fi
\providecommand{\bibinfo}[2]{#2}
\providecommand{\eprint}[2][]{\url{#2}}

\bibitem[{\citenamefont{Clarke and Braginski}(2004)}]{ClarkeSQUID04}
\bibinfo{author}{\bibfnamefont{J.}~\bibnamefont{Clarke}} \bibnamefont{and}
  \bibinfo{author}{\bibfnamefont{A.~I.} \bibnamefont{Braginski}},
  \emph{\bibinfo{title}{The SQUID Handbook}}, vol. \bibinfo{volume}{1,2}
  (\bibinfo{publisher}{Wiley-VCH}, \bibinfo{address}{Weinheim},
  \bibinfo{year}{2004}).
  
\bibitem[{\citenamefont{Simmonds et~al.}(2001)\citenamefont{Simmonds, 
Marchenkov, Hoskinson, Davis, and Packard}}]{SimmondsQuantumN01}
\bibinfo{author}{\bibfnamefont{R.~W.} \bibnamefont{Simmonds \textit{et al.}}},
   \bibinfo{journal}{Nature} \textbf{\bibinfo{volume}{412}}, \bibinfo{pages}{55}
  (\bibinfo{year}{2001}).
  
\bibitem[{\citenamefont{Hoskinson et~al.}(2001)\citenamefont{Hoskinson, Sato, and Packard}}]{HoskinsonSuperfluidPRB06}
\bibinfo{author}{\bibfnamefont{E.}~\bibnamefont{Hoskinson}},
\bibinfo{author}{\bibfnamefont{Y.}~\bibnamefont{Sato}},
\bibinfo{author}{\bibfnamefont{R.}~\bibnamefont{Packard}},
  \bibinfo{journal}{Phys. Rev. B} \textbf{\bibinfo{volume}{74}},
  \bibinfo{pages}{100509} (\bibinfo{year}{2006}).
  
\bibitem{atomtronics}
\bibinfo{author}{\bibfnamefont{B.~T.} \bibnamefont{Seaman \textit{et al.}}},
  \bibinfo{journal}{Phys. Rev. A}
  \textbf{\bibinfo{volume}{75}}, \bibinfo{pages}{023615}
  (\bibinfo{year}{2007}).
\bibinfo{author}{\bibfnamefont{R.~A.} \bibnamefont{Pepino  \textit{et al.}}},
  \bibinfo{journal}{Phys. Rev. Lett.}
  \textbf{\bibinfo{volume}{103}}, \bibinfo{pages}{140405}
  (\bibinfo{year}{2009}).
\bibinfo{author}{\bibfnamefont{A.} \bibnamefont{Ruschhaupt}},
  \bibnamefont{and} 
  \bibinfo{author}{\bibfnamefont{J.~G.} \bibnamefont{Muga}},
  \bibinfo{journal}{Phys. Rev. A} \textbf{\bibinfo{volume}{70}},
  \bibinfo{pages}{061604} (\bibinfo{year}{2004}).
\bibinfo{author}{\bibfnamefont{J.~J.} \bibnamefont{Thorn \textit{et al.}}},
  \bibinfo{journal}{Phys. Rev. Lett.} \textbf{\bibinfo{volume}{100}},
  \bibinfo{pages}{240407} (\bibinfo{year}{2008}).
\bibinfo{author}{\bibfnamefont{J.~A.} \bibnamefont{Stickney}},
  \bibinfo{author}{\bibfnamefont{D.~Z.} \bibnamefont{Anderson}},
  \bibnamefont{and} \bibinfo{author}{\bibfnamefont{A.~A.}
  \bibnamefont{Zozulya}}, \bibinfo{journal}{Phys. Rev. A}
  \textbf{\bibinfo{volume}{75}}, \bibinfo{pages}{013608}
  (\bibinfo{year}{2007}).

\bibitem[{\citenamefont{Likharev}(1979)\citenamefont{Likharev}}]{LikharevRMP79}
\bibinfo{author}{\bibfnamefont{K.~K.} \bibnamefont{Likharev}},
   \bibinfo{journal}{Rev. Mod. Phys.} \textbf{\bibinfo{volume}{51}}, \bibinfo{pages}{101}
  (\bibinfo{year}{1979}).
  
\bibitem[{\citenamefont{Davis and Packard}(2002)\citenamefont{Davis and Packard}}]{DavisRMP02}
\bibinfo{author}{\bibfnamefont{J.~C.} \bibnamefont{Davis} \bibnamefont{and}},
\bibinfo{author}{\bibfnamefont{R.~E.} \bibnamefont{Packard}},
\bibinfo{journal}{Rev. Mod. Phys.} \textbf{\bibinfo{volume}{74}}, \bibinfo{pages}{741}
  (\bibinfo{year}{2001}).

\bibitem[{\citenamefont{Albiez et~al.}(2005)\citenamefont{Albiez, Gati,
  F\"olling, Hunsmann, Cristiani, and Oberthaler}}]{AlbiezDirectPRL05}
\bibinfo{author}{\bibfnamefont{M.}~\bibnamefont{Albiez \textit{et al.}}},
\bibinfo{journal}{Phys. Rev. Lett.}
  \textbf{\bibinfo{volume}{95}}, \bibinfo{pages}{010402}
  (\bibinfo{year}{2005}).

\bibitem[{\citenamefont{Levy et~al.}(2007)\citenamefont{Levy, Lahoud, Shomroni,
  and Steinhauer}}]{LevyACDCN07}
\bibinfo{author}{\bibfnamefont{S.}~\bibnamefont{Levy \textit{et al.}}},
\bibinfo{journal}{Nature} \textbf{\bibinfo{volume}{449}},
\bibinfo{pages}{579} (\bibinfo{year}{2007}).

\bibitem[{\citenamefont{Landau}(1941)}]{LandauJPUSSR41}
\bibinfo{author}{\bibfnamefont{L.~D.} \bibnamefont{Landau}},
  \bibinfo{journal}{J. Phys. (USSR)} \textbf{\bibinfo{volume}{5}},
  \bibinfo{pages}{71} (\bibinfo{year}{1941}).

\bibitem[{end()}]{endnote}
\bibinfo{note}{This assumes that the spatial scale of any perturbing potential
  is much less than the healing length}.

\bibitem[{\citenamefont{Pitaevskii and
  Stringari}(2003)}]{PitaevskiiBose-Einstein03}
\bibinfo{author}{\bibfnamefont{L.~P.} \bibnamefont{Pitaevskii}}
  \bibnamefont{and}
  \bibinfo{author}{\bibfnamefont{S.}~\bibnamefont{Stringari}},
  \emph{\bibinfo{title}{Bose-Einstein Condensation}}
  (\bibinfo{publisher}{Clarendon}, \bibinfo{address}{Oxford},
  \bibinfo{year}{2003}).

\bibitem[{\citenamefont{Barenghi et~al.}(2001)\citenamefont{Barenghi, Donnelly,
  , and Vinen}}]{BarenghiQuantized01}
\bibinfo{editor}{\bibfnamefont{C.~F.} \bibnamefont{Barenghi}},
  \bibinfo{editor}{\bibfnamefont{R.~J.} \bibnamefont{Donnelly}}, ,
  \bibnamefont{and} \bibinfo{editor}{\bibfnamefont{W.~F.} \bibnamefont{Vinen}},
  eds., \emph{\bibinfo{title}{Quantized Vortex Dynamics and Superfluid
  Turbulence}} (\bibinfo{publisher}{Springer-Verlag},
  \bibinfo{address}{Berlin}, \bibinfo{year}{2001}).

\bibitem[{\citenamefont{Feynman}(1955)}]{FeynmanApplicationPLTP55}
\bibinfo{author}{\bibfnamefont{R.}~\bibnamefont{Feynman}},
  \bibinfo{journal}{Prog. Low Temp. Phys.} \textbf{\bibinfo{volume}{1}},
  \bibinfo{pages}{17} (\bibinfo{year}{1955}).

\bibitem[{\citenamefont{Avenel and Varoquaux}(1985)}]{AvenelObservationPRL85}
\bibinfo{author}{\bibfnamefont{O.}~\bibnamefont{Avenel}} \bibnamefont{and}
  \bibinfo{author}{\bibfnamefont{E.}~\bibnamefont{Varoquaux}},
  \bibinfo{journal}{Phys. Rev. Lett.} \textbf{\bibinfo{volume}{55}},
  \bibinfo{pages}{2704} (\bibinfo{year}{1985}).

\bibitem[{\citenamefont{Amar et~al.}(1992)\citenamefont{Amar, Sasaki, Lozes,
  Davis, and Packard}}]{AmarQuantizedPRL92}
\bibinfo{author}{\bibfnamefont{A.}~\bibnamefont{Amar \textit{et al.}}},
  \bibinfo{journal}{Phys. Rev. Lett.} \textbf{\bibinfo{volume}{68}},
  \bibinfo{pages}{2624} (\bibinfo{year}{1992}).

\bibitem[{\citenamefont{Huebener}(2001)}]{HuebenerMagnetic01}
\bibinfo{author}{\bibfnamefont{R.~P.} \bibnamefont{Huebener}},
  \emph{\bibinfo{title}{Magnetic Flux Structures in Superconductors}}
  (\bibinfo{publisher}{Springer}, \bibinfo{year}{2001}).

\bibitem[{\citenamefont{Neely et~al.}(2010)\citenamefont{Neely, Samson,
  Bradley, Davis, and Anderson}}]{NeelyObservationPRL10}
\bibinfo{author}{\bibfnamefont{T.~W.} \bibnamefont{Neely \textit{et al.}}},
  \bibinfo{journal}{Phys. Rev. Lett.} \textbf{\bibinfo{volume}{104}},
  \bibinfo{pages}{160401} (\bibinfo{year}{2010}).

\bibitem[{\citenamefont{Onofrio et~al.}(2000)\citenamefont{Onofrio, Raman,
  Vogels, Abo-Shaeer, Chikkatur, and Ketterle}}]{OnofrioObservationPRL00}
\bibinfo{author}{\bibfnamefont{R.}~\bibnamefont{Onofrio \textit{et al.}}},
  \bibinfo{journal}{Phys. Rev. Lett.} \textbf{\bibinfo{volume}{85}},
  \bibinfo{pages}{2228} (\bibinfo{year}{2000}).

\bibitem[{\citenamefont{Engels and Atherton}(2007)}]{EngelsStationaryPRL07}
\bibinfo{author}{\bibfnamefont{P.}~\bibnamefont{Engels}} \bibnamefont{and}
  \bibinfo{author}{\bibfnamefont{C.}~\bibnamefont{Atherton}},
  \bibinfo{journal}{Phys. Rev. Lett.} \textbf{\bibinfo{volume}{99}},
  \bibinfo{pages}{160405} (\bibinfo{year}{2007}).

\bibitem[{\citenamefont{Ryu et~al.}(2007)\citenamefont{Ryu, Andersen,
  Clad\'{e}, Natarajan, Helmerson, and Phillips}}]{RyuObservationPRL07}
\bibinfo{author}{\bibfnamefont{C.}~\bibnamefont{Ryu \textit{et al.}}},
\bibinfo{journal}{Phys. Rev. Lett.}
  \textbf{\bibinfo{volume}{99}}, \bibinfo{pages}{260401}
  (\bibinfo{year}{2007}).

\bibitem[{ywas()}]{yeswearesure}
\bibinfo{note}{We apply a strong barrier during evaporation, removing it adiabatically when well below the condensation temperature, several seconds before we create circulation.}

\bibitem[{\citenamefont{Wright et~al.}(2008)\citenamefont{Wright, Leslie, and
  Bigelow}}]{WrightOpticalPRA08}
\bibinfo{author}{\bibfnamefont{K.~C.} \bibnamefont{Wright}},
  \bibinfo{author}{\bibfnamefont{L.~S.} \bibnamefont{Leslie}},
  \bibnamefont{and} \bibinfo{author}{\bibfnamefont{N.~P.}
  \bibnamefont{Bigelow}}, \bibinfo{journal}{Phys. Rev. A}
  \textbf{\bibinfo{volume}{77}}, \bibinfo{pages}{041601}
  (\bibinfo{year}{2008}).

\bibitem[{\citenamefont{Andersen et~al.}(2006)\citenamefont{Andersen, Ryu,
  Clad\'e, Natarajan, Vaziri, Helmerson, and
  Phillips}}]{AndersenQuantizedPRL06}
\bibinfo{author}{\bibfnamefont{M.~F.} \bibnamefont{Andersen \textit{et al.}}},
\bibinfo{journal}{Phys. Rev. Lett.}
  \textbf{\bibinfo{volume}{97}}, \bibinfo{pages}{170406}
  (\bibinfo{year}{2006}).

\bibitem[{\citenamefont{Madison et~al.}(2000)\citenamefont{Madison, Chevy,
  Wohlleben, and Dalibard}}]{MadisonVortexPRL00}
\bibinfo{author}{\bibfnamefont{K.~W.} \bibnamefont{Madison \textit{et al.}}},
  \bibinfo{journal}{Phys. Rev. Lett.} \textbf{\bibinfo{volume}{84}},
  \bibinfo{pages}{806} (\bibinfo{year}{2000}).

\bibitem[{\citenamefont{Freilich et~al.}(2010)\citenamefont{Freilich, Bianchi,
  Kaufman, Langin, and Hall}}]{FreilichReal-TimeS10}
\bibinfo{author}{\bibfnamefont{D.~V.} \bibnamefont{Freilich \textit{et al.}}},
  \bibinfo{journal}{Science} \textbf{\bibinfo{volume}{329}},
  \bibinfo{pages}{1182} (\bibinfo{year}{2010}).

\bibitem[{\citenamefont{Ramanathan et~al.}()\citenamefont{Ramanathan}}]{RamanathanTBP}
  \bibinfo{author}{\bibfnamefont{A.}~\bibnamefont{Ramanathan}~\bibnamefont{\textit{et al.}}},
  \bibinfo{note}{to be published}.
  
\bibitem[{unc()}]{uncertaintynote}
  \bibinfo{note}{Uncertainties herein are the uncorrelated combination
  of 1$\sigma$ statistical and systematic uncertainties unless stated otherwise}. 

\bibitem[{\citenamefont{Watanabe et~al.}(2009)\citenamefont{Watanabe, Dalfovo,
  Piazza, Pitaevskii, and Stringari}}]{WatanabeCriticalPRA09}
\bibinfo{author}{\bibfnamefont{G.}~\bibnamefont{Watanabe \textit{et al.}}},
  \bibinfo{journal}{Phys. Rev. A} \textbf{\bibinfo{volume}{80}},
  \bibinfo{pages}{053602} (\bibinfo{year}{2009}).

\bibitem[{\citenamefont{Stringari}(1998)}]{StringariDynamicsPRA98}
\bibinfo{author}{\bibfnamefont{S.}~\bibnamefont{Stringari}},
  \bibinfo{journal}{Phys. Rev. A} \textbf{\bibinfo{volume}{58}},
  \bibinfo{pages}{2385} (\bibinfo{year}{1998}).

\bibitem[{\citenamefont{Piazza et~al.}(2009)\citenamefont{Piazza, Collins, and
  Smerzi}}]{PiazzaVortex-inducedPRA09}
\bibinfo{author}{\bibfnamefont{F.}~\bibnamefont{Piazza}},
  \bibinfo{author}{\bibfnamefont{L.~A.} \bibnamefont{Collins}},
  \bibnamefont{and} \bibinfo{author}{\bibfnamefont{A.}~\bibnamefont{Smerzi}},
  \bibinfo{journal}{Phys. Rev. A} \textbf{\bibinfo{volume}{80}},
  \bibinfo{pages}{021601} (\bibinfo{year}{2009}).

\end{thebibliography}
\end{document}